\documentclass[aps,showpacs,showkeys,groupedaddress,twocolumn]{revtex4}

\usepackage{epsfig}

\usepackage{amscd}
\usepackage{amsmath}
\usepackage{bm}
\usepackage[dvips]{color}                 
\usepackage{graphicx}
\usepackage{eufrak}
\usepackage{eucal}

\newcommand{\beq}{\begin{equation}}
\newcommand{\eeq}{\end{equation}}
\newcommand{\bea}{\begin{eqnarray}}
\newcommand{\eea}{\end{eqnarray}}

\newcommand{\benn}{\begin{displaymath}}
\newcommand{\eenn}{\end{displaymath}}

\def\[{\left[}
\def\]{\right]}

\begin{document}

\title{\bf  Temperature Effects in a Fermi Gas with Population Imbalance}

\author{Heron Caldas\footnote{{\tt hcaldas@ufsj.edu.br}}, and A. L. Mota\footnote{{\tt motaal@ufsj.edu.br}}}

\affiliation{Universidade Federal de S\~{a}o Jo\~{a}o del Rei, S\~{a}o Jo\~{a}o del Rei, 36301-160, MG, Brazil}

\begin{abstract}
We investigate temperature effects in a Fermi gas with imbalanced spin populations. From the general expression of the thermal gap equation we find, in {\it weak coupling limit}, an analytical expression for the transition temperature $T_c$ as a function of various possibilities of chemical potential and mass asymmetries between the two particle species. For a range of asymmetry between certain specific values, this equation always has two solutions for $T_c$ which has been interpreted as a reentrant phenomena or a pairing induced by temperature effect. We show that the lower $T_c$ is never related to a stable solution. The same results are obtained in {\it strong coupling limit}. The thermodynamical potential is carefully analyzed to avoid the consideration of the unstable solutions. We also obtain the tricritical points for the chemical potential and mass imbalanced cases, and beyond these points we properly minimize the thermodynamic potential to find the stable and metastable first order transition lines.

\end{abstract}

\maketitle
\bigskip

The recent advances in experiments with ultracold fermionic atoms have provided the possibility for the understanding of superfluidity in several physical situations, from high temperature superconductivity to the pairing of quarks in the cores of neutron stars. 

When a two fermion species system have the same number of spin-up and spin-down particles, its ground state is described by the well known Bardeen-Cooper-Schrieffer (BCS) theory of superconductivity~\cite{BCS}. In this case, if the temperature is below a certain critical temperature ($T_c$), fermions with opposite spins interact near their common Fermi surface resulting in pair formation, even at arbitrarily weak coupling. For temperatures above $T_c$ the system is found to be in the normal state i.e., the fermions are unpaired. 

The possibility of pairing formation in imbalanced systems, where the chemical potentials or the numbers of fermionic species are different, was first pointed out more than 4 decades ago by Sarma~\cite{Sarma}. Since then, various exotic phases have been proposed for the mismatched case, such as the Larkin, Ovchinnikov, Fulde and Ferrel (LOFF)-phase~\cite{Larkin}, the breached pair superfluid phase (BP)~\cite{Liu}, deformed Fermi surfaces~\cite{Sedrakian}, phase separation in real space~\cite{PRL1,Heron}, and also the possibility of new coexisting phases in the BEC regime~\cite{Dan}. Other pairing mechanisms beyond BCS, such as P-wave superfluidity, have also been investigated~\cite{Aurel}. Observation of phase separation between a fully paired superfluid core surrounded by the unpaired excess atoms, have been reported independently by the Rice~\cite{Hulet,Hulet2} and MIT~\cite{Ketterle,Ketterle2} groups. This phase separation can be viewed in terms of three phase transitions of different nature. Differently from the standard BCS thermodynamical phase transition we mentioned above, in an imbalanced system the phase transitions that may happen are: {\bf I.} At zero temperature ($T$), increasing the chemical potential or number asymmetry the system undergoes a (first-order) quantum phase transition to the normal state~\cite{PRL1}; {\bf II.} Still at $T=0$, first order phase transitions occur from the normal to a phase separation (PS) phase, and from PS to a spatially homogeneous (magnetized) superfluid as the interaction parameter $1/k_Fa$ is varied~\cite{Dan,Parish}, with a tricritical point sitting in the fully polarized line~\cite{Parish}, and {\bf III.} Lowering the temperature, a system with different and fixed number particles phase separates into a unpolarized superfluid core surrounded by a polarized normal shell, provided the asymmetry does not exceed a critical value~\cite{Ketterle,Ketterle2}. Thus, it is of crucial importance to address the issue of the temperature effects in imbalanced Fermi gases~\cite{Ketterle3}. Some recent works have investigated the relevant problem of imbalanced Fermi gases at finite temperature~\cite{Yi,He1,various}. However, there are only few recent theoretical studies exploring, concomitantly, the temperature effects and the stability of population and mass imbalanced Fermi gases~\cite{Sa,Wu,Paananen1,Parish}.

In this work we investigate thermodynamical phase transition in a two species homogeneous system. We carry out a quantitative study of the lowest energy state of an imbalanced Fermi gas on the BCS side of the resonance. In this regime unexpected temperature effects manifest, such as the appearance of two solutions for the mean-field (MF) transition temperature $0<T_{c,1}<T_{c,2}$. A gap $\Delta$ would emerge at $T_{c,1}$, increase up to a maximum value, and then decrease, vanishing at $T_{c,2}$. A premature interpretation of this nonmonotonic behavior for $\Delta$ and its two critical temperatures is that temperature favors pairing~\cite{He1}. On the contrary, we find that temperature (heat, to be more precise) always acts in the direction of disrupting the fermion pairs, mainly in imbalanced systems, at least within the model under consideration. We show systematically, employing MF theory, both at weak and strong coupling, that the lower critical temperature is related to an unstable solution of the gap equation, and thus does not correspond to a true lowest energy (stable) state. However, we point out that the reentrant phenomena could manifest when the BEC-BCS crossover theory is considered. In this case, fluctuation contributions (necessary for the introduction of the concept of a {\it pseudogap}) are taken into account~\cite{Levin1,Levin2}. 

More important than only identifying the unphysical solutions of the gap equation, we demonstrate: (i) Analytically we derive, in the weak coupling limit, an expression for the second order phase transition line from which we found the tricritical points $(\delta \mu_c,T_c)$ and $(\delta m_c,T_c)$. (ii) Numerically, we obtain the physical second- and first-order phase transition lines separating the superfluid and normal phases, both in the weak and strong coupling limits. (iii) We find the metastable first order curves in the phase diagrams $T$ vs. $\delta \mu$, and $T$ vs. $\delta m$, ignored in all previous analysis on temperature effects in mass imbalanced Fermi gases.

{\it The Model}. To begin with, let us consider a nonrelativistic dilute (i.e., the particles interact through a short-range attractive interaction) cold fermionic system, described by the following Hamiltonian

\begin{eqnarray}
\label{eq-1}
{\cal H}&=&H-\sum_{k,\alpha}\mu_{\alpha}n_{\alpha}\\
\nonumber
&=&\sum_{k} {\epsilon}^{a}_k a^{\dagger}_{k} a_{k}+{\epsilon}^{b}_k b^{\dagger}_{k} b_{k}
- g \sum_{k,k'} a^{\dagger}_{k'} b^{\dagger}_{-k'} 
b_{-k} a_{k},\,
\end{eqnarray}
where $a^{\dagger}_{k}$, $a_k$ are the creation and annihilation operators for the $a$ particles (and the same for the $b$ particles)
and ${\epsilon}^{\alpha}_k$ are their dispersion relation, defined by ${\epsilon}^{\alpha}_k=\xi_k^{\alpha}-\mu_{\alpha}$,
with $\xi_k^{\alpha}=\frac{k^2}{2m_{\alpha}}$ and $\mu_{\alpha}$ being the chemical potential of the (non-interacting) $\alpha$-species, $\alpha=a,b$. To reflect an attractive (s-wave) interaction between particles $a$ and $b$ we take $-g<0$.

{\it The Thermodynamic Potential and the Thermal Gap Equation}. From the Hamiltonian (\ref{eq-1}) one obtains the grand potential from which all thermodynamical quantities of interest can be obtained:

\begin{figure}[t]
\includegraphics[height=2.4in]{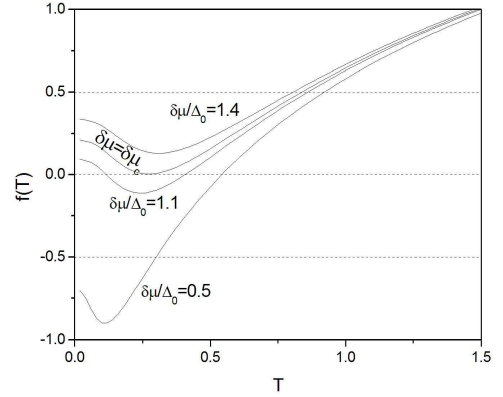}
\caption{The function $f(T/\Delta_0)$ for different chemical potential asymmetries. For small asymmetries $\delta \mu < \delta \mu_0$, there is only one solution for $T_c$. For $\delta \mu < \delta \mu < \delta \mu_c$ there are two solutions for the transition temperature. When $\delta \mu > \delta \mu_c$ there are no solutions for $T_c$ from the gap equation.}
\label{fig1}
\end{figure}

\begin{eqnarray}
\label{tp}
\Omega=\frac{\Delta^2}{g} +  \sum_{k} \Big[ \epsilon_k^{+}-E_k-T \ln(e^{-\beta {\cal{E}}_k^{a}}+1)  \\
\nonumber
-T \ln(e^{-\beta {\cal{E}}_k^{b}}+1)\Big],
\end{eqnarray}
where ${\cal{E}}_k^{a,b}= E_k \pm \epsilon_k^{-}$ are the quasiparticle excitations, with $E_k=\sqrt{ {\epsilon_k^{+}}^2+\Delta^2 }$, and $\epsilon_k^{\pm} \equiv \frac {\epsilon_k^a \pm \epsilon_k^b}{2}$. Deriving the equation above with respect to $\Delta$, we obtain the gap equation

\begin{equation}
\label{ap9}
\frac{1}{g}= \sum_{k} \frac{1}{2 E_k} \left( 1-f_{k} - g_{k} \right),
\end{equation}
where $f_{k}$ and $g_{k}$ are the Fermi distribution functions $f_{k},g_{k}=1/(e^{\beta {\cal{E}}_k^{a,b}}+1)$, with $\beta=1/T$, where we have set the Boltzmann constant equal to one. The critical temperature $T_c$ is, by definition, the temperature at which $\Delta=0$. Then Eq.~(\ref{ap9}) becomes

\begin{equation}
\label{ap10}
\frac{1}{g}= \sum_{k} \frac{1}{\epsilon_k^a+\epsilon_k^b} \left(1-\frac{1}{e^{\beta_c \epsilon_k^a}} - \frac{1}{e^{\beta_c \epsilon_k^b}} \right),
\end{equation}
where $\beta_c=1/T_c$. 

{\it Analytical Solutions}. In the weak coupling limit, it is possible to obtain a compact expression for $T_c$ from the equation above. After the momentum integration, Eq.~(\ref{ap10}) can be written as~\cite{Heron3}

\beq
\label{tc2}
T_c=\frac{\sigma\Delta_0}{2 \pi} e^{-\frac{1}{2} {\cal F}(a_c)},
\eeq
where $\sigma \equiv \frac{M}{\sqrt{m_a m_b}}$ is a dimensionless parameter reflecting the mass asymmetry, $M=\frac{m_a m_b}{m_a + m_b}$ is the reduced mass, $\Delta_0$ is the $T=0$ BCS gap parameter in the weak coupling limit, ${\cal F} (x) \equiv \Psi (\frac{1}{2}+\frac{ix}{\pi })+\Psi (\frac{1}{2}-\frac{ix}{\pi })$, with $\Psi$ being the digamma function, defined as $\Psi(z)=\frac{\Gamma'(z)}{\Gamma(z)}$, where z is a complex number with a positive real component, $\Gamma$ is the gamma function, $\Gamma'$ its derivative, and $a_c \equiv \frac{\beta_c}{2} \eta = \frac{\beta_c}{2} \frac{m_b \mu_b- m_a \mu_a}{m_a + m_b}$. Equation~(\ref{tc2}) gives $T_c$ as a function of the mass and chemical potential asymmetry, encoded in $ \eta$. Due to the highly non-linear term ${\cal F} (a_c \neq 0)$ in this equation, it is possible to find analytical solutions for $T_c$ only when the Fermi momentum of the two particle species match, $P_F^a=P_F^b$, where $P_F^{\alpha}=\sqrt{2m_{\alpha}\mu_{\alpha}}$. This can happen when $m_a=m_b$ and $\mu_a=\mu_b$, which results in ${\cal F}(a_c=0)=-2\gamma - 4 \ln(2)$, and $\sigma=1/2$, yielding the standard BCS result $T_c/\Delta_0=\frac{e^{\gamma}}{\pi}$ or, in a very exotic situation, where $m_a \neq m_b$ and $\mu_a \neq \mu_b$ but, in spite of the asymmetries, the particles have the same Fermi surface, $m_a \mu_a=m_b \mu_b$, giving $T_c/\Delta_0=2\sigma \frac{e^{\gamma}}{\pi}$~\cite{Heron3,Heron4}. As can be seen from the expression for $a_c$, the critical temperature $T_c$ given by Eq.~(\ref{tc2}) is a maximum when the Fermi surfaces for the $a$ and $b$ particles are equal, since ${\cal F}(a_c=0)$ is a minimum. We obtain numerically the critical temperature of an imbalanced system, not restricted to the previously analyzed equal Fermi surfaces cases. First we note that Eq.~(\ref{tc2}) can be written in terms of dimensionless quantities as

\begin{equation}
ln\Big( \frac{\sigma}{2\pi} \frac{1}{T_c/\Delta_0} \Big)=\frac{1}{2} {\cal F}\Big( \frac{1}{4T_c/\Delta_0} \Big( \frac{\delta m}%
{\tilde{m}} \frac{\mu}{\Delta_0} + \frac{\delta\mu}{\Delta_0} \Big) \Big), 
\label{coeff}
\end{equation}
where $\delta m = m_b-m_a$, $\tilde{m}=m_b+m_a$, $\mu = \mu_a + \mu_b$, $\delta \mu=\mu_b-\mu_a$. Since we also consider mass asymmetry, it is convenient to rewrite $\sigma$ and $M$ in terms of the nondimensional mass ratio $-1< \frac{\delta m}{\tilde{m}} < 1$, as $\sigma=\frac{1}{2}\sqrt{1-\left( \frac{\delta m}{\tilde{m}} \right)^2}$, and $M=\frac{1}{2} m_a \left(1+ \frac{\delta m}{\tilde{m}} \right)$. As the chemical potential difference $\delta \mu$ increases, Eq.~(\ref{coeff}) has two solutions, i.e., there are two temperatures at which $\Delta=0$. This can be seen in Fig.~(\ref{fig1}), where we plot
 
\begin{equation}
f(T)=\frac{1}{2}{\cal F}\Big( \frac{1}{4T/\Delta_0} \Big( \frac{\delta m}%
{\tilde{m}} \frac{\mu}{\Delta_0} + \frac{\delta\mu}{\Delta_0} \Big) \Big)-ln\Big( \frac{\sigma}{2\pi} \frac{1}{T/\Delta_0} \Big), 
\label{f(T)}
\end{equation}
for different values of $\delta \mu/\Delta_0$, with equal masses, as a function of $T$. Throughout the paper we use, when necessary, $\mu/\Delta_0=7$. We can see that, for $\delta \mu/\Delta_0$ small enough there is only one solution of Eq.~(\ref{coeff}). At a certain value of the chemical potential asymmetry $\delta \mu_0$, Eq.~(\ref{coeff}) starts to develop two solutions, $T_{c,1}$ and $T_{c,2}$. We find, as we shall see below, that $T_{c,2}$ corresponds to a second order phase transition temperature, whereas $T_{c,1}$ corresponds to the temperature at which the local maximum of the thermodynamic potential disappears. Increasing the chemical potential asymmetry beyond $\delta \mu_0$ a critical value is reached $\delta \mu_c$, after which the system presents a first order phase transition at a critical temperature $T_c$. Finally, at a higher chemical potential asymmetry
$\delta\mu_{CC}/\Delta_0=\sqrt{2}$, the Chandrasekhar-Clogston limit of fermionic superfluidity~\cite{Chandresakar,Clogston}, the system is a normal Fermi liquid, at any temperature, even $T=0$.

\begin{figure}[t]
\includegraphics[height=2.4in]{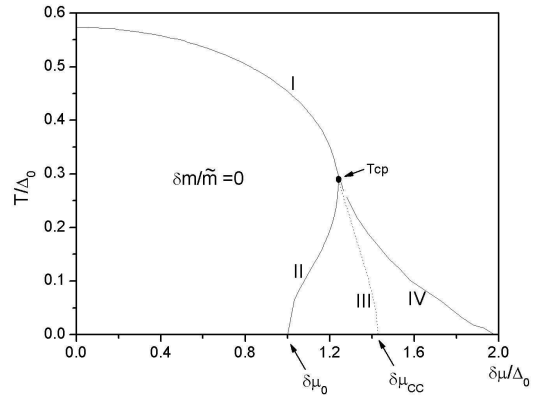}
\caption{The critical temperature as a function of the chemical potential asymmetry, plotted from Eq.~(\ref{tc2}), in the weak coupling approximation.}
\label{fig2}
\end{figure}

The critical temperature as a function of $\delta \mu$ is presented in Fig.~(\ref{fig2}) for equal masses. Curves ${\rm I}$ and ${\rm III}$ are the second and first order phase transition lines, respectively. Curves ${\rm II}$ and ${\rm IV}$ are the unstable and metastable, second and first order, respectively, phase transition lines. Below the tricritical point one has to properly minimize the free energy rather than using the gap equation as the transition becomes first order. Thus, the second order transition line comes simply from the solution of the gap equation. To construct the first order line we numerically search for the temperature $T_c$ at which $\Omega(\Delta_{min}, T_c)= \Omega(\Delta=0, T_c)$, where $\Delta_{min}$ is the non-trivial minimum of $\Omega$.

\begin{figure}[t]
\includegraphics[height=2.4in]{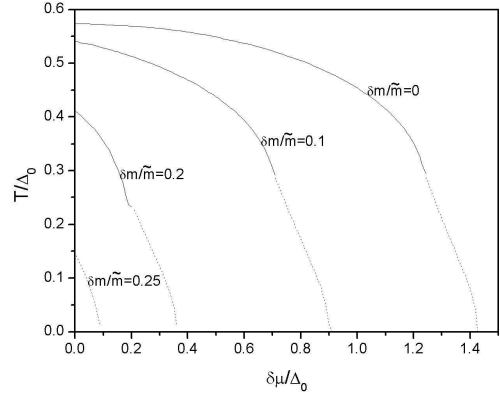}
\caption{The critical temperature as a function of $\delta \mu/\Delta_0$ for several mass ratios $\delta m/\tilde{m}$, as indicated in the figure, with $\mu/\Delta_0=2$.}
\label{fig3}
\end{figure}

Along curves ${\rm II}$ and  ${\rm IV}$ $\Delta=0$, but does not correspond to the absolute minimum of $\Omega$. Thus the true stable phases of an imbalanced fermionic system are below curves ${\rm I}$ and ${\rm III}$ (the superfluid phase) and above them (the normal phase). The tricritical point $(\delta \mu_c,T_c)=(1.2308\Delta_0,0.2713\Delta_0)$, and the chemical potential differences $\delta \mu_0$ and $\delta \mu_{CC}$ are shown in the figure. The small bumps at line ${\rm IV}$ are due to a not high enough numerical accuracy, but do not affect the qualitative behavior of this curve. In Fig.~(\ref{fig3}) we show the critical temperature as a function of $\delta \mu$ for different (positive) mass ratios ($\delta m/\tilde{m}$ runs from 0 (outer curve) to 0.25 (inner curve), with $\mu/\Delta_0=2$).

We can derive analytical expressions for $\delta \mu_0$ and $\delta \mu_c$ even when $\delta m \neq 0$. Looking at Fig.~(\ref{fig1}) we see that the chemical potential asymmetry at which Eq.~(\ref{coeff}) begins to present two solutions, $\delta\mu_0$, occurs when $f(T=0)=0$. If we observe that, as $T \rightarrow 0$, ${\cal F}(a_c)$ goes exactly to $2 ln(|a_c|/\pi)$, then Eq.~(\ref{coeff}) can
be rewritten as $ln\Big( \frac{\sigma}{2\pi} \frac{1}{T_c/\Delta_0} \Big)=\frac{1}{2} 2 ln\Big( \big| \frac{1}{\pi} \frac{1}{4T_c/\Delta_0} \Big( \frac{\delta m}%
{\tilde{m}} \frac{\mu}{\Delta_0} + \frac{\delta\mu}{\Delta_0} \Big) \big| \Big)$. This implies in

\begin{equation}
\frac{\delta \mu_0}{\Delta_0} = \pm 2 \sigma - \frac{\delta m}{\tilde{m}} \frac{\mu}{\Delta_0}.
\end{equation}
For $\delta m=0$ ($\sigma=1/2$) the two critical temperatures start to show up at $\frac{\delta \mu_0}{\Delta_0} = \pm 1$, exactly as obtained numerically in Fig.~(\ref{fig2}) for $\delta \mu >0$. Conversely, for fixed chemical potential imbalance and (artificially) varying mass asymmetry, the two $T_c$ will begin to appear at 

\begin{equation}
\frac{\delta m_0}{\tilde{m}}=  \frac{-\frac{\mu \delta \mu}{\Delta_0^2} \pm \sqrt{1+ \left(\frac{\mu}{\Delta_0}\right)^2-\left(\frac{\delta \mu}{\Delta_0} \right)^2}}{1+\left(\frac{\mu}{\Delta_0} \right)^2}.
\end{equation} 
The behavior of $T_c/\Delta_0$ as a function of $\frac{\delta m}{\tilde{m}}$ is similar to that presented in Figs.~(\ref{fig2}) and (\ref{fig7}). The tricritical point $(\delta \mu_c, T_c)$ is obtained by imposing the vanishing of the quadratic ($\equiv A$) and the quartic ($\equiv B$) terms in the Landau expansion of the free energy, Eq.~(\ref{tp}). The coefficients $A$ and $B$ are given by $f(T)$ and $\kappa df(T)/dT$, respectively, where $\kappa$ is a constant~\cite{WP}. Graphically this can be seen in Fig.~(\ref{fig1}) for $\delta\mu=\delta\mu_c$, with $\delta \mu_c$ corresponding to the chemical potential asymmetry where $T_c$ occurs at the minimum of $f(T)$ and above which there are no more solutions for $f(T)=0$. From Eq.~(\ref{f(T)}) we obtain

\begin{equation}
\frac{df(T)}{dT}=-\frac{1}{T}+\frac{a}{2T}{\cal F}^\prime (a)=0. 
\label{dfdt}
\end{equation}
We can rewrite $a \equiv \frac{\beta}{2} \eta = \frac{\beta}{2} \frac{m_b \mu_b- m_a \mu_a}{m_a + m_b}$ as $a=\frac{\Delta_0}{4T}\Big( \frac{\delta m}{\tilde{m}} \frac{\mu}{\Delta_0} + \frac{\delta\mu}{\Delta_0} \Big)$. The solutions of Eq.~(\ref{dfdt}), $a{\cal F}^\prime(a)=2$, are $\pm a_0$, where $a_0 \equiv 1.1343$, i.e., $a_0$ is the argument that minimizes $f(T)$. Thus, the temperature that minimizes $f(T)$ is $T_{c,min}=\frac{\Delta_0}{4a_0}\Big(\big| \frac{\delta m}{\tilde{m}} \frac{\mu}{\Delta_0} + \frac{\delta\mu}{\Delta_0} \big| \Big)$. As discussed above, the (tri-)critical chemical potential $\delta\mu_c$ is determined by $T_{c,min}=T_c$ in Eq.~(\ref{tc2}), $\frac{\Delta_0}{4a_0}\Big( \frac{\delta m}{\tilde{m}} \frac{\mu}{\Delta_0} + \frac{\delta\mu_c}{\Delta_0} \Big) =T_{c,min}= \frac{\sigma\Delta_0}{2\pi}e^{-\frac{1}{2}{\cal F}(a_0)}$, resulting in

\begin{equation}
\frac{\delta\mu_c}{\Delta_0}= \frac{\pm 2a_0\sigma}{\pi}e^{-\frac{1}{2}{\cal F}(a_0)}-\frac{\delta m}{\tilde{m}}%
\frac{\mu}{\Delta_0}, 
\label{deltamuc}
\end{equation}
where $\sigma=\sigma(\frac{\delta m}{\tilde{m}})$. From the equation above one clearly sees that a positive mass imbalance reduces the critical chemical potential imbalance. Computing $\frac{\delta\mu_c}{\Delta_0}$ for $\delta m=0$ ($e^{-\frac{1}{2}{\cal F}(a_0)}\approx 3.4089$), we find $\frac{\delta\mu_c}{\Delta_0}\approx \pm 1.2308$.
Also, when $\delta m=0$, we can use Eq.~(\ref{tc2}) and replace $\frac{\sigma}{\pi}e^{-\frac{1}{2}{\cal F}(a_0)}=2T_c/\Delta_0$ in Eq.(\ref{deltamuc}), obtaining
\begin{equation}
\frac{\delta\mu_c}{\Delta_0}= \pm 4a_0 \frac{T_c}{\Delta_0},
\end{equation}
and the critical temperature for the equal masses system at the critical chemical potential results $\frac{T_c}{\Delta_0} \approx 0.2713$, which compares exactly with those numerically obtained in Fig.~(\ref{fig2}). This result should be contrasted to that of the standard (symmetric) BCS result $\frac{T_c}{\Delta_0}=\frac{e^{\gamma}}{\pi}\approx 0.5669$. The point $(\delta \mu_c,T_c)$ is a tricritical point, with a line of first-order transition emerging, and hitting the $\delta \mu$ axis at $(\delta \mu_{CC},0)$. Note that the phase diagram of the equal mass system is symmetric with respect to the $T_c/\Delta_0$ axis. From Eq.~(\ref{deltamuc}), one can see that for $\frac{\delta m}{\tilde{m}}>\frac{\delta m_{i}}{\tilde{m}}$ ($\frac{\delta m}{\tilde{m}} < - \frac{\delta m_{i}}{\tilde{m}}$), where $\frac{\delta m_{i}}{\tilde{m}} \equiv \frac{\lambda}{\sqrt{\lambda^2+\left( \mu / \Delta_0 \right)^2}}$, and $\lambda \equiv \frac{a_0}{\pi}e^{-\frac{1}{2}{\cal F}(a_0)}\approx1.2312$, the phase diagram can be entirely shifted to the left (right) of the $T_c/\Delta_0$ axis, exhibiting two negative (positive) values for the tricritical chemical potential imbalance, which we denote $\delta \mu_{c1}/\Delta_0$, and $\delta \mu_{c2}/\Delta_0$. Beyond these tricritical points there are also generally the CC points limiting superfluidity between $(\delta \mu_{CC1}/\Delta_0,0)$ and $(\delta \mu_{CC2}/\Delta_0,0)$. As we will discuss later, in trapped systems the chemical potentials of the two species vary with the position from the center to the edge of the trap. This can assure the variation of $\delta \mu/\Delta_0$ necessary to the formation of shell structures~\cite{Lin,Parish,Pao2,Paananen2}, composed of a superfluid phase ($\delta \mu_{CC1}/\Delta_0 < \delta \mu/\Delta_0 <\delta \mu_{CC2}/\Delta_0$) sandwiched between a normal core ($\delta \mu/\Delta_0 < \delta \mu_{CC1}/\Delta_0$) and an outer normal phase ($\delta \mu/\Delta_0 > \delta \mu_{CC2}/\Delta_0$), in real space. Thus we see that only mass imbalanced trapped systems allow shell structure configuration. Physically, one can understand the displacement of the (weak coupling) phase diagram analysing the position of the top of the dome: the maximum critical temperature corresponds, for a given mass asymmetry,
to the chemical potential difference at which the Fermi surfaces match. It is an easy task
to show that the maximum $T_c$ in Eq.~(\ref{tc2}) occurs when $\frac{\delta m}{\tilde{m}}=-\frac{\delta\mu}{\mu}$ or, equivalently, $m_a \mu_a=m_b \mu_b$ (see also Eq.~(\ref{coeff})). Note from this expression
that a higher positive mass asymmetry implies in a more negative chemical potential asymmetry at the top of the dome.

We can proceed in the same way as in the case of varying $\delta \mu$, and find the (tri-)critical mass imbalance

\begin{equation}
\frac{\delta m_c}{\tilde{m}}=  \frac{-\frac{\mu \delta \mu}{\Delta_0^2} \pm \lambda \sqrt{\lambda^2 + \left(\frac{\mu}{\Delta_0}\right)^2-\left(\frac{\delta \mu}{\Delta_0} \right)^2}}{\lambda^2+\left(\frac{\mu}{\Delta_0} \right)^2},
\end{equation} 
with $\delta \mu$ fixed. Equivalently, when $\delta \mu /\Delta_0=0$ the phase diagram $T_c/\Delta_0$ vs. $\delta m / \tilde{m}$ is symmetric with respect to the $\delta m / \tilde{m}=0$ axis, and the dome is shifted to the left or to the right depending on the sign of a nonvanishing $\delta \mu /\Delta_0$.

\begin{figure}[t]
\includegraphics[height=2.4in]{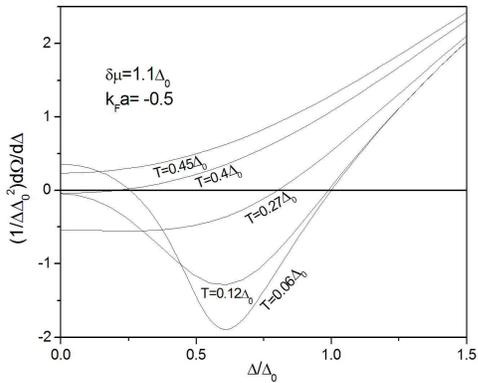}
\caption{The plot of $G \equiv \frac{1}{\Delta_0^2}\frac{1}{2\Delta} \frac{d \Omega}{d \Delta}$ as a function of $\Delta/\Delta_0$ for various temperatures. The zeros of $G$ give the gaps that are candidates for the minimum of $\Omega$. For $T$ greater than $\approx .41 \Delta_0$ there are no more solutions for $G$, which means that this is the actual critical temperature of the system (see Fig.~(\ref{fig2})).}
\label{fig4}
\end{figure}

\begin{figure}[t]
\includegraphics[height=2.4in]{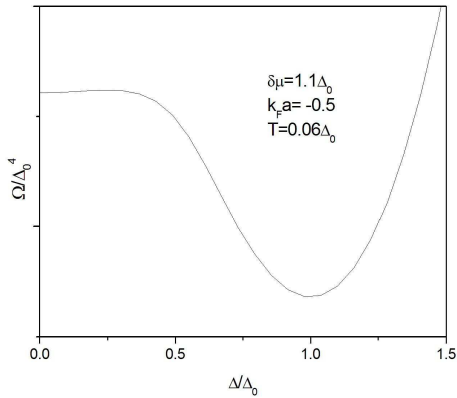}
\caption{$\Omega/\Delta_0^4$ as a function of $\Delta/\Delta_0^4$ for $k_Fa=-0.5$, $\delta \mu =1.1 \Delta_0$, and $T=0.06\Delta_0$. For these parameters the gap equation has two solutions, as shown in Fig.~(\ref{fig4}): the smaller gap is related to a local maximum of $\Omega$, and the larger gap corresponds to a minimum of $\Omega$.}
\label{fig5}
\end{figure}

{\it Numerical Solutions}. {\bf a.} Weak coupling. To find the gap which correspond to stable solutions of the system, as well as the first order, second order and the metastable phase transitions, we numerically implement the thermodynamic potential expression, Eq.~(\ref{tp}), and its derivative with respect to $\Delta$, and solve them for several masses and chemical potential asymmetries, at different temperatures. The derivative of $\Omega$ is proportional to $2 \Delta$, so $\Delta=0$ is always a extremum of $\Omega$. To focus on the $\Delta \neq 0$ solution of $d \Omega/d \Delta=0$, we introduce the (nondimensional) function $G \equiv \frac{1}{2\Delta_0^2}\frac{1}{\Delta} \frac{d \Omega}{d \Delta}$. Thus, the zeros of $G$ give the non-trivial extrema of $\Omega$. We plot, in Fig.~(\ref{fig4}), $G$ as a function of $\Delta/\Delta_0$. The coupling $g$ can be related to the two-body s-wave scattering length $a$ by $\frac{1}{g}=-\frac{M}{2 \pi a} + \int\frac{d^3 k}{(2 \pi)^3} \frac{1}{2 \tilde{\xi_k}}$, where $\tilde{\xi_k}=\frac{\vec{k}^2}{4M}$. Then we use this expression to trade $g$ for the dimensionless parameter $k_Fa$ and, doing that, we can establish contact with experiment. The zeros of $G$ for a given set of parameters $\mu$, $\delta \mu$, $M$, $T$, and $k_Fa$ give the gaps that are candidates for the minimum of $\Omega$. To avoid the consideration of unstable solutions~\cite{Yi}, we also verify the behavior of $\Omega$ versus $\Delta$ with these same parameters. We take a $\delta \mu > \delta \mu_0$ and a $T<T_{c,1}$ since the system would be in the normal phase with these parameters (see Fig.~(\ref{fig2}), that has been plotted from Eq.~(\ref{tc2})). We can observe that there are two gaps for $\delta \mu =1.1 \Delta_0$, and $T=0.06\Delta_0$. The smaller gap ($\equiv\Delta_1$) does not correspond to a minimum of $\Omega$, it is a local maximum, whereas the larger gap ($\equiv\Delta_2$) is related to the (superfluid) stable solution, as can be seen in Fig.~(\ref{fig5}) where we plot $\Omega/\Delta_0^4$ as a function of $\Delta/\Delta_0$. Still in Fig.~(\ref{fig4}), we see that as the temperature is increased from $T=0.06\Delta_0$ up to $T = 0.12\Delta_0$ which is $\approx T_{c,1}$ the system ceases to have two $T_c$. This is an indication that the lower $T_c$ of Fig.~(\ref{fig2}) (remember, the two $T_c$ appear only for $\delta \mu_0 < \delta \mu < \delta \mu_c$) is the critical temperature of the unstable gap. Another way of seeing that the smaller gap corresponds to an unstable phase is the observation of the quasiparticle excitations depicted in Fig.~(\ref{fig6}). With the smaller gap the quasiparticles of the lower branch would have negative excitations, exhibiting the same unstable zero temperature behavior~\cite{PRL1,Heron}. For $T_{c,1} < T < T_{c,2}$ there is only one solution, which always corresponds to a stable phase. For even greater temperatures, or $T > T_{c,2}$ there are no more solutions for the (second order transition) gap equation, Eq.~(\ref{ap10}). For $\delta\mu > \delta\mu_c$ we compare the thermodynamical potential evaluated at $\Delta=0$ with the same quantity evaluated at the solutions $\Delta_0(T)$ of the equation $G=0$, which gives the extrema of the thermodynamic potential. The temperature where $\Omega(\Delta=0,T)=\Omega(\Delta_0(T),T)$ corresponds to the first order phase transition critical temperature $T_c$. For temperatures greater than $T_c$, Eq. $G=0$ can still present solutions, but they do not correspond to the absolute minimum of $\Omega$. Curves IV (associated with metastable phases) in figures (\ref{fig2}) and (\ref{fig7}) correspond to the temperatures above which $G=0$ has no more solutions. 

\begin{figure}[t]
\includegraphics[height=2.4in]{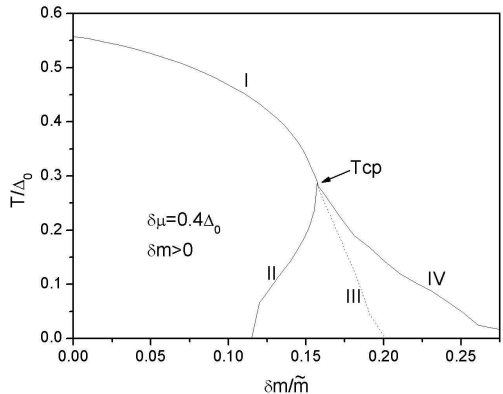}
\caption{The critical temperature as a function of the (positive) masses asymmetry for $\delta\mu = 0.4\Delta_0$, in the weak coupling limit.}
\label{fig5.5}
\end{figure}

Fig.~({\ref{fig5.5}}) shows the results for the dependence of the critical temperatures with the mass asymmetry. 
As anticipated, the behavior is quite similar to that presented for the chemical potential asymmetric case, as discussed above. Nevertheless, we must remark that the dependence of the critical temperature $T_c$ with mass asymmetry is not the same as for the chemical potential asymmetric case, as can be viewed from Eq. (\ref{coeff}), since $\sigma$ depends on $\frac{\delta m}{\tilde{m}}$ but not on $\delta\mu$. In obtaining this figure, we kept $m_a$ constant ($=\Delta_0$) and varied the mass of particle $b$. In experiments, it would be equivalent to measure the critical temperatures of the system with species $a$ fixed, for several $b$ particles with different masses, but always keeping the chemical potential asymmetry fixed. Above some specific mass asymmetry, $\delta m_0/\tilde{m}$, the system presents two critical temperatures (as happens for varying $\delta \mu$, with fixed $\delta m$), one corresponding to the temperature where a local maximum of $\Omega$ disappears (line II), and the other to the critical temperature where the system undergoes a second order phase transition (line I). The behavior of the system as a function of the mass asymmetry also possesses a tricritical point $Tcp$, where the second order phase transition line (I) meets the first order phase transition line (III) at $\delta m_c/\tilde{m}$, as we already discussed. There is also a critical mass asymmetry, above which the system no longer presents a superfluid phase, even at $T=0$. This would be the equivalent of the Chandrasekhar-Clogston (CC) limit of superfluidity for the case of mass imbalance. Besides, for temperatures below those of curve IV, the system could still present a (metastable) local minimum of $\Omega$.

\begin{figure}[t]
\includegraphics[height=2.4in]{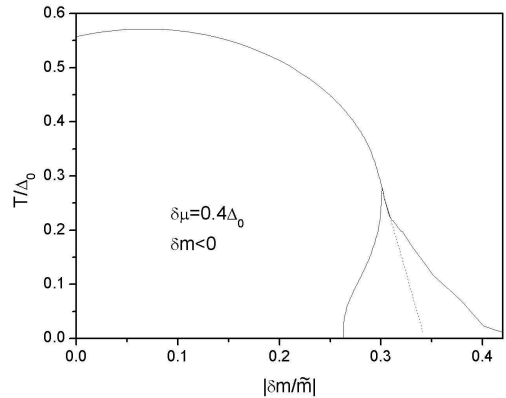}
\caption{The critical temperature as a function of the (negative) masses asymmetry for $\delta\mu = 0.4\Delta_0$, in the weak coupling limit. The critical temperature presents its maximum value at a non vanishing mass asymmetry. This phase diagram is the continuation of that of Fig.~(\ref{fig5.5}) for the negative values of $\delta m/\tilde{m}$. The higher $T_c$ is located in the top of the dome when the condition $\frac{\delta m}{\tilde{m}}=-\frac{\delta\mu}{\mu}$ is met, i.e., when $P_F^a=P_F^b$.}
\label{fig5.6}
\end{figure}

\begin{figure}[t]
\includegraphics[height=2.4in]{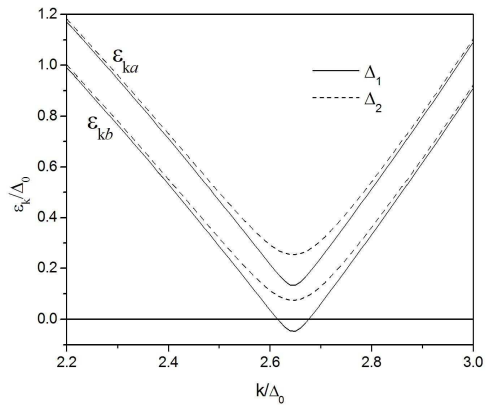}
\caption{The quasiparticles excitation energies behavior as a function of the two gaps $\Delta_1<\Delta_2$ found for $T=0.06\Delta_0$. With the smaller gap, the lower branch has quasiparticles (${\cal{E}}_{k,b}$) with negative energy.}
\label{fig6}
\end{figure}

\begin{figure}[t]
\includegraphics[height=2.4in]{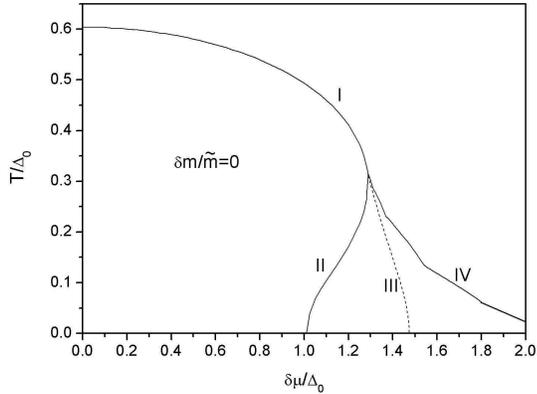}
\caption{$T_c/\Delta_0$ as a function of $\delta \mu / \Delta_0$ for $\delta m=0$ from a numerical evaluation of Eq.~(\ref{ap10}). This curve has the same qualitative behavior of that shown in Fig.~(\ref{fig2}), obtained in the weak coupling limit, but clearly shows that $\delta \mu_c(k_Fa=-10)>\delta \mu_{c}(k_Fa=-0.5)$ .}
\label{fig7}
\end{figure}

It is interesting to note that the possibility of pairing of particles with different masses and chemical potentials results in situations not present in the purely chemical potential ($\delta m=0$) or purely mass ($\delta \mu=0$) asymmetric cases. If we have, for example, $m_a > m_b$ (which implies in a negative $\frac{\delta m}{\tilde{m}}$) in considering $\mu_b > \mu_a$, the phase diagram behaves as depicted in Fig (\ref{fig5.6}), i.e., the critical temperature is higher for a non vanishing mass asymmetry. We can observe it, for instance, in the analytical solution for the weak coupling regime, as a consequence of the fact that $T_c$ given by Eq.(\ref{coeff}) does not have its maximum value for $\delta m/\tilde{m} = 0$, but rather when the Fermi surfaces match.

{\bf b.} Moderate to strong coupling. Now we investigate the solutions of Eq.~(\ref{ap10}) in the moderate to strong coupling limit, $k_Fa=-10$, and show the resulting $T_c/\Delta_0$ as a function of $\delta \mu / \Delta_0$ in Fig.~(\ref{fig7}). As one can sees, this curve has the same qualitative behavior of that in the weak coupling limit. We obtain next, the solutions of the gap equation, the thermodynamic potential and the quasi particle excitations, which are shown in Figs.~(\ref{fig8}) to (\ref{fig10}) below, respectively. As seen in the weak-coupling regime (Fig.~(\ref{fig6})), for some temperatures ($T < T_{c,1}$, where $T_{c,1}$ is the temperature at which the unstable gap vanishes, for any $k_Fa$) the quasiparticles of the lower branch have negative excitations (Fig.~(\ref{fig10})) between two momentum values, and consequently, two Fermi surfaces with gapless modes (BP2). This is in the heart of the instability of the breached-pair or phase separation in momentum space phases both at zero~\cite{PRL1} and finite temperature, as we have shown here. However, it is worth to mention that a state where the lower branch excitation crosses zero once, defining only one Fermi surface (BP1), is found to be stable~\cite{Pao,Yi,Parish}.

\begin{figure}[t]
\includegraphics[height=2.4in]{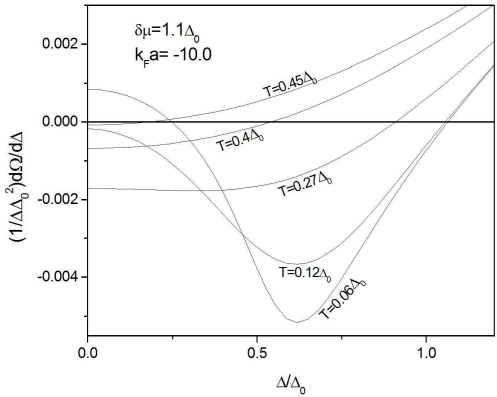}
\caption{The plot of $G \equiv \frac{1}{\Delta_0^2}\frac{1}{2\Delta} \frac{d \Omega}{d \Delta}$ as a function of $\Delta/\Delta_0$ for various temperatures. The zeros of $G$ give the gaps that are candidates for the minimum of $\Omega$. The $T_c$ in the strong coupling limit, $\approx .44 \Delta_0$ is larger than that of the weak-coupling limit, for same $\delta \mu$, (see Fig.~(\ref{fig4})).}
\label{fig8}
\end{figure}

To take into account the trap effects, usually one employs the local density approximation (LDA) via $\mu_{\alpha} \to \mu(\vec{r})_{\alpha}=\mu_{\alpha}-V(\vec{r})$, where $\mu_{\alpha}$ are the (global) chemical potentials introduced before, and $V(\vec{r})$ is the trapping potential. As we mentioned already, trapped fermions with unequal masses can exhibit a shell structure. It is very interesting to note that our MF weak-coupling analytical results for $\delta m=0$ in Eq.~(\ref{coeff}), are independent of $\mu(\vec{r})=\mu-2V(\vec{r})$, and depends only on $\delta \mu(\vec{r})=\delta \mu$. Since the shell structure manifests only for trapped fermions with unequal masses~\cite{Lin,Parish,Pao2}, this is an indication that the dependence of the main structure of the zero and finite temperature phase diagram on $\delta m ~\mu(\vec{r})$ is maintained in the strong coupling limit.

\begin{figure}[t]
\includegraphics[height=2.4in]{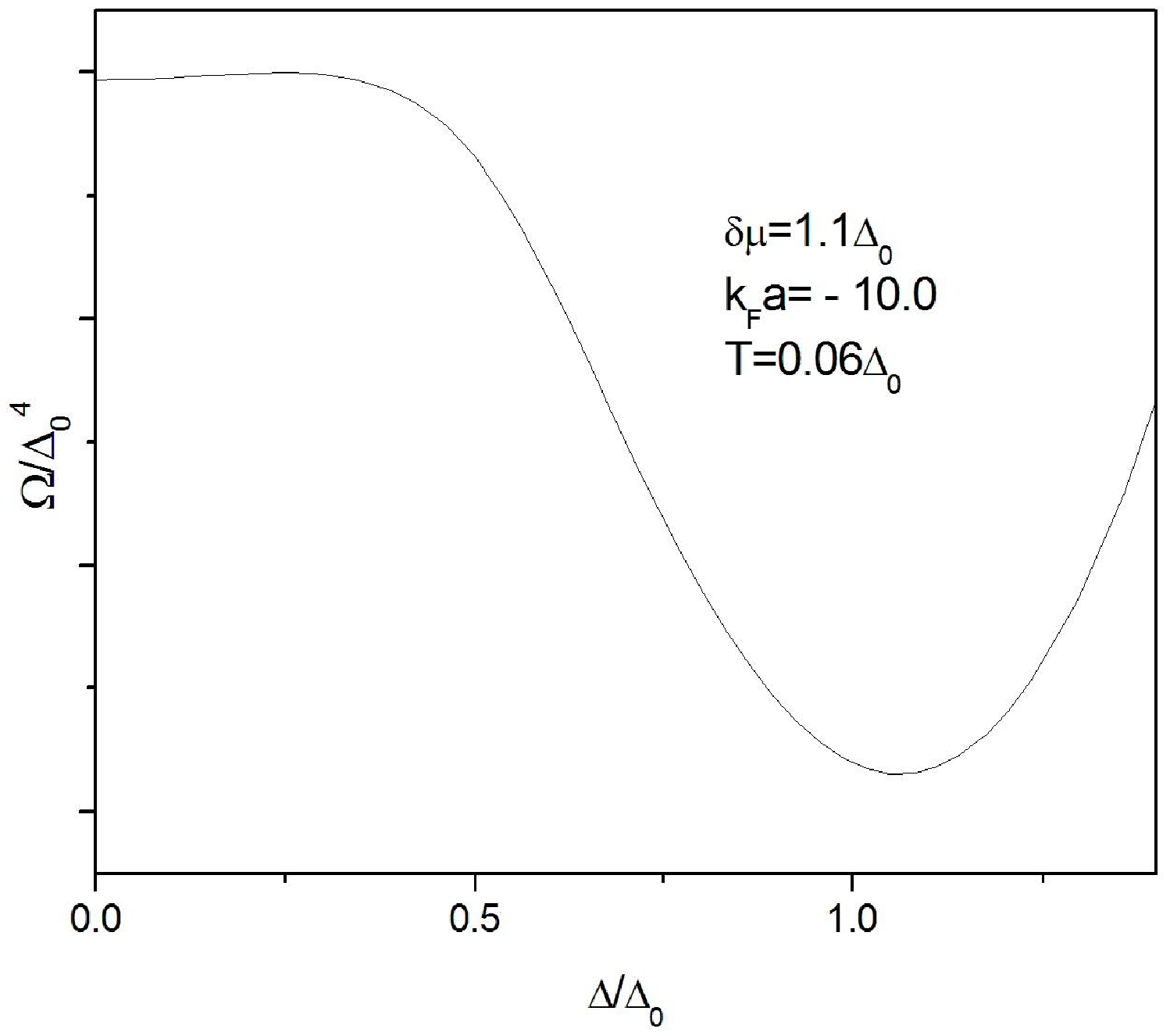}
\caption{$\Omega/\Delta_0^4$ as a function of $\Delta/\Delta_0^4$ for $k_Fa=-10$, $\delta \mu =1.1 \Delta_0$, and $T=0.06\Delta_0$. As observed in the weak coupling regime (see Fig.~(\ref{fig5})), for these parameters the gap equation has two solutions, as shown in Fig.~(\ref{fig8}): the smaller gap is related to a local maximum of $\Omega$, and the larger gap corresponds to a minimum of $\Omega$.}
\label{fig9}
\end{figure}

{\it Discussion and Conclusion}. We have investigated temperature effects and thermodynamical phase transitions in fermionic gases composed by two particle species whose Fermi surfaces or densities do not match. We have observed two of the three phase transitions we mentioned in the introduction. At zero temperature (and at low fixed $T$, as in Figs.~(\ref{fig5}) and (\ref{fig9})), increasing the chemical potential asymmetry, at $\delta \mu_c$ the minimum of $\Omega/\Delta_0^4$ jumps from $\Delta/\Delta_0=1$ ($\Delta/\Delta_0=1.1$ in Fig.~(\ref{fig9})) to $\Delta/\Delta_0=0$. At fixed $\delta \mu < \delta \mu_c$, increasing $T$ the minimum of $\Omega/\Delta_0^4$ goes smoothly from some $\Delta/\Delta_0 < 1$ to $\Delta/\Delta_0=0$. For $\delta \mu_c < \delta \mu < \delta \mu_{CC}$, increasing $T$ the minimum of $\Omega/\Delta_0^4$ jumps from some $\Delta/\Delta_0 < 1$ to $\Delta/\Delta_0=0$. We observed the same conclusions for the varying mass situation. Besides the second and first order lines, respectively, associated with the phase transitions we just mentioned, we have found the unstable and metastable lines related with these phase transitions. We have shown that for specific values of the chemical potential (and/or mass) asymmetry the thermal gap equation of imbalanced systems has two solutions. The smaller gap, corresponding to the BP phase, always represents unstable solutions. We have found for the first time, in the weak coupling limit, an analytical expression for the tricritical chemical potential difference, which in turn depends on the mass asymmetry. A positive (negative) mass asymmetry reduces (increases) the positive tricritical chemical potential imbalance, as a consequence of the displacement of the dome to the negative (positive) side of the $T_c/\Delta_0$ axis. These conclusions were obtained also in the strong coupling regime. We have also shown that raising the interaction parameter $k_Fa$, results in a small increasing of the (critical) chemical potential differences $\delta \mu_c$ and $\delta \mu_{CC}$ and, equivalently, of the critical mass differences. We also have identified the presence of metastable states beyond the first-order transition lines (curves III in Figs. (\ref{fig2}), (\ref{fig5.5}), (\ref{fig5.6}), and (\ref{fig7})) and the temperatures at which these metastable solutions disappear in a first order phase transition (curves IV in the same figures). Given the current possibility of controlling several parameters in ultracold fermionic systems, such as the interaction strength, densities, trapping potential, and temperature, it would be possible the experimental preparation of long-lived metastable states of imbalanced superfluids, allowing the study of thermodynamic and decay process of these states, a subject of interest in several areas of physics~\cite{VC,MC,GK}. We hope that our results for the imbalanced masses can be verified, considering the limitations of the MF theory, in the new generation of experiments that could in principle be set up with two kinds of fermionic atoms with opposite spins and different masses~\cite{Martin2}.

\begin{figure}[t]
\includegraphics[height=2.4in]{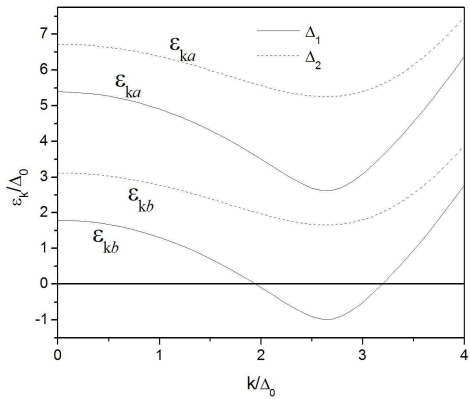}
\caption{The quasiparticles excitation energies behavior as a function of the two gaps $\Delta_1<\Delta_2$ found for for $k_Fa=-10$, and $T=0.06\Delta_0$. As in the weak coupling limit (Fig.~6), with the smaller gap, the lower branch has quasiparticles (${\cal{E}}_{k,b}$) with negative energy. However, in the strong coupling regime the range of momenta at which the particles have negative energy (meaning that they are single) is bigger than in the weak coupling limit.}
\label{fig10}
\end{figure}

\vspace{1cm}

\label{conc}

\section*{Acknowledgments}
We thank D. Sheehy for helpful discussions. H. Caldas acknowledges partial support by the Brazilian agencies CNPq and FAPEMIG.



\end{document}